\documentclass[12pt]{article}
 \usepackage{epsfig}
 \usepackage {graphicx}
 \usepackage[latin1]{inputenc}
 \usepackage{amsfonts}
 \usepackage{amsthm}
 \usepackage{amssymb, latexsym, amsmath}
 \newcommand{\inc}{{\it i}}

 \newcommand{\be}{\begin{equation}}
 \newcommand{\ee}{\end{equation}}
 \newcommand{\ba}{\begin{eqnarray}}
 \newcommand{\ea}{\end{eqnarray}}

 \newcommand{\erbold}{\mbox{{\boldmath $\vec r$}}}

 \newcommand{\epsilonbold}{\mbox{{\boldmath $\epsilon$}}}

  \newcommand{\eRbold}{\mbox{{\boldmath $\vec{R}$}}}
  
  \newcommand{\Vbold}{\mbox{\boldmath $\vec {\boldmath{\,V}}$}}

\begin{document}
 \title{
                 ${{~~~~~~~~~~~~~~~~~~}^{^{^{
                  }}}}$\\
                  ~\\
 {\Large{\textbf{Can the tidal quality factors of terrestrial planets and moons scale as positive powers of the tidal frequency?\\
 ~\\
 ~\\}
            }}}
 \author{
 {\Large{Michael Efroimsky}}\\
 {\small{US Naval Observatory, Washington DC 20392 USA}}\\
 {\small{e-mail: ~michael.efroimsky @ usno.navy.mil~}}\\ }
 \date{}

 \maketitle
 \begin{abstract}
 In geophysics and seismology, it is a common knowledge that the quality factors $Q$ of the mantle and crust materials scale as the tidal frequency to
 a positive fractional power (Karato 2007, Efroimsky \& Lainey 2007). In astronomy, there exists an equally common belief that such rheological models
 introduce discontinuities into the equations and thus are unrealistic at low frequencies. We demonstrate that, while such models indeed make the
 conventional expressions for the tidal torque diverge for vanishing frequencies, the emerging infinities reveal not the impossible nature of
 one or another rheology, but a subtle flaw in the underlying mathematical model of friction. Flawed is the common misassumption that the tidal force
 and torque are inversely proportional to the quality factor. In reality, they are proportional to the sine of the tidal phase lag, 
 while the inverse quality factor is commonly identified with the tangent of the lag. The sine and tangent of the lag are close 
 everywhere {\it{except in the vicinity of the zero frequency}}. Reinstating of this detail tames the fake
 infinities and rehabilitates the ``impossible" scaling law (which happens to be the actual law the mantles obey).

 This preprint is a pilot paper. A more comprehensive treatise on tidal torques is to be published (Efroimsky \& Williams 2009).

 \end{abstract}

\pagebreak

 \section{Introduction.}

 We are considering bodily tides in a primary perturbed by a secondary. Each elementary volume of the primary is subject to a tide-raising potential,
 which in general is not periodic but can be expanded into a sum of periodic terms. We shall assume that the primary is homogeneous and incompressible.
 Although simplistic, the model provides a good qualitative understanding of tidal evolution of both the primary's spin and the secondary's orbit

 \section{Linearity of the tide}

 \subsection{Two aspects of linearity}

 We assert deformations to be linear. Each tidal harmonic $\,W_{\it{l}}\,$ of the
 potential disturbance produced by the secondary generates a linear deformation of
 the primary's shape, while each such deformation amends the potential of the
 primary with an addition proportional to the Love number $\,k_{\it{l}}\,$.
 Linearity also implies that the energy attenuation rate $\langle \stackrel{\centerdot}{E}(\chi)
 \rangle$ at each frequency $\,\chi\,$ depends solely on the value of $\,\chi\,$
 and on the amplitude $\,E_{peak}(\chi)\,$, and is not influenced by the other
 harmonics. Thus,
 \ba
 \langle\,\dot{E}(\chi)\,\rangle \;=\;-\;\frac{\textstyle\chi E_{peak}(\chi)}{\textstyle Q(\chi)}\;
 \label{gg}\label{L11}
 \ea
 or, equivalently:
 \ba
 \Delta E_{cycle}(\chi)\;=\;-\;\frac{2\;\pi\;E_{peak}(\chi)}{Q(\chi)}\;\;\;,
 \label{dd}\label{L12}
 \ea
 $\Delta E_{cycle}(\chi)\,$ being the one-cycle energy loss. The so-defined quality
 factor $\,Q(\chi)\,$ corresponding to some frequency $\,\chi\,$ is interconnected
 with the phase lag $\,\epsilon(\chi)\,$ corresponding to the same
 frequency.

  If $\,E_{peak}(\chi)\,$ in (\ref{L11} - \ref{L12}) is agreed to denote the peak {\it{energy}} stored at frequency $\,\chi\,$,
 the appropriate $Q$ factor is connected to the phase lag $\,\epsilon(\chi)\,$ through
 \ba
 Q^{-1}_{\textstyle{_{energy}}}~=~\tan|\epsilon|~~~.
 \label{DI6_1}
 \label{DI6}
 \ea
 If $E_{peak}(\chi)$ is defined as the peak {\it{work}},
  the corresponding $Q$ factor is related to the lag via
 \ba
 Q^{-1}_{\textstyle{_{work}}}\;=\;\frac{\tan |\epsilon|}{1\;-\;\left(\;\frac{\textstyle \pi}{\textstyle 2}\;-\;
 |\epsilon|\;\right)  \;\tan|\epsilon|}\;\;\;,~~~~~
 \label{A81}
 \label{L14}
 \ea
 as demonstrated in the Appendix below.

 Both definitions render a vanishing $Q$ for the lag approaching $\,\pi/2\,$, and both result in the same approximation for $Q$ in
 the limit of a small lag:
 \ba
 Q^{-1}_{\textstyle{_{energy}}}~=~\sin|\epsilon|\;+\;O(\epsilon^2)\;=\;|\epsilon|\;+\;O(\epsilon^2)\;\;\;.
 \label{DI6_2}
 \ea
 Hence definitions (\ref{DI6} - \ref{L14}) make $\,1/Q\,$ a good approximation to $\,\sin\epsilon\,$
 for small lags only.

 This makes the essence of the standard, linear theory of bodily tides. The model
 permits for the freedom of choice of the functional dependency of the quality factor
 upon the tidal frequency. Whatever the form of this dependence, the basic idea of the theory is
 the following: the tide-raising potential is expanded over periodic terms, whereafter
 (a) the material's response is assumed to be linear, and (b) the overall attenuation
 rate is asserted to be a sum of rates corresponding to the involved frequencies.

 \subsection{Goldreich's admonition:\\ a general difficulty stemming from nonlinearity}

 Introduced empirically as a means to figleaf our lack of knowledge of the attenuation process in its full complexity, the notion of $\,Q\,$ has proven
 to be practical due to its smooth and universal dependence upon the frequency and temperature. At the same time, this empirical treatment has its
 predicaments and limitations. Its major inborn defect was brought to light by Peter Goldreich who pointed out that the attenuation rate at a
 particular frequency depends not only upon the appropriate Fourier component of the stress, but also upon the {\emph{overall}} stress. This happens
 because for real minerals each quality factor $\,Q(\chi_{\textstyle{_i}})\,$ bears dependence not only on the frequency $\,\chi_{\textstyle{_i}}\,$
 itself, but also on the magnitude of the $\,\chi_{\textstyle{_i}}-$component of the stress and, most importantly, also on the {\emph{overall}} stress.
 This, often-neglected, manifestation of nonlinearity may be tolerated only when the amplitudes of different harmonics of stress are comparable.
 However, when the amplitude of the principal mode is orders of magnitude higher than that of the harmonics (tides being the case), then the principal
 mode will, through this nonlinearity, make questionable our entire ability to decompose the overall attenuation into a sum over frequencies. Stated
 differently, the quality factors corresponding to the weak harmonics will no longer be well defined physical parameters.

 Here follows a quotation from Goldreich (1963):\\
 {\emph{``... Darwin and Jeffreys both wrote the tide-raising potential
 as the sum of periodic potentials. They then proceeded to consider the
 response of the planet to each of the potentials separately. At first
 glance this might seem proper since the tidal strains are very small
 and should add linearly. The stumbling block in this procedure, however,
is the amplitude dependence of the specific dissipation function. In the case
of the Earth, it has been shown by direct measurement that $\,Q\,$ varies by
an order of magnitude if we compare the tide of frequency  $\,2\omega - 2n \,$
with the tides of frequencies $\, 2\omega - n \,$,  $\, 2\omega - 3n \,$, and
$\,\frac{\textstyle 3}{\textstyle 2}n\,$. This is because these latter tides
have amplitudes which are smaller than the principle tide (of frequency $\,
2\omega - 2n \,$ ) by a factor of eccentricity or about 0.05. It may still
appear that we can allow for this amplitude dependence of Q merely by adopting
an amplitude dependence for the phase lags of the different tides.
Unfortunately, this is really not sufficient since a tide of small amplitude
will have a phase lag which increases when its peak is reinforcing the peak of
the tide of the major amplitude. This non-linear behaviour cannot be treated
in detail since very little is known about the response of the planets to
tidal forces, except for the Earth."}}

 On these grounds, Goldreich concluded the paragraph with an important
 warning that we {\emph{``use the language of linear tidal theory, but we
 must keep in mind that our numbers are really only parametric fits to a
 non-linear problem."}}

 In order to mark the line beyond which this caveat cannot be ignored, let
 us first of all recall that the linear approximation remains applicable
 insofar as the strains do not approach the nonlinearity threshold, which
 for most minerals is of order $\,10^{-6}\,$. On approach to that
 threshold, the quality factors may become dependent upon the strain
 magnitude. In other words, in an attempt to extend the expansion
 (\ref{gg} - \ref{dd}) to the nonlinear case, we shall have to introduce,
 instead of $\,Q(\chi_{\textstyle{_i}})\,$, some new functions
 $\,Q(\,\chi_{\textstyle{_i}}\,,\;E_{_{peak}}(\chi_{\textstyle{_i}})\,,
 \;E_{_{overall}}\,)\;$. (Another complication is that in the nonlinear
 regime new frequencies will be generated, but we shall not go there.)
 Now consider a superposition of two forcing stresses -- one at
 the frequency $\,\chi_{\textstyle{_1}}\,$ and another at
 $\,\chi_{\textstyle{_2}}\,$.
 Let the amplitude $\;E_{_{peak}}(\chi_{\textstyle{_1}})\;$ be close or
 above the nonlinearity threshold, and
 $\;E_{_{peak}}(\chi_{\textstyle{_2}})\;$ be by an order or two
 of magnitude smaller than $\;E_{_{peak}}(\chi_{\textstyle{_1}})\;$.
 To adapt the linear machinery (\ref{gg} - \ref{dd})  to the
 nonlinear situation, we have to write it as
 \ba
 \langle\,\dot{E}~\rangle~=~\langle\,\dot{E}_1~\rangle~+~\langle\,\dot{E}_2
 ~\rangle~=~-~\chi_{\textstyle{_1}}~\frac{~E_{_{peak}}(\chi_{\textstyle{_1}
 })~}{Q(\,\chi_{\textstyle{_1}}\,,~E_{_{peak}}(\chi_{\textstyle{_1}})\,)}~
 -~\chi_{\textstyle{_2}}~\frac{\,E_{_{peak}}(\chi_{\textstyle{_2}})\,}{Q(
 \chi_{\textstyle{_2}}\,,\;E_{_{peak}}(\chi_{\textstyle{_1}})\,,\;
 E_{_{peak}}(\chi_{\textstyle{_2}})\,)}\;\;\;,\;\;\;\;\;
 \label{409}
 \ea
 the second quality factor bearing a dependence not only upon the frequency
 $\,\chi{\textstyle{_2}}\,$ and the appropriate magnitude $\;E_{_{peak}}(
 \chi_{\textstyle{_2}})\;$, but also upon the magnitude of the
 {\emph{first}} mode, $\;E_{_{peak}}(\chi_{\textstyle{_1}})\;$, -- this
 happens because it is the first mode which makes a leading contribution
 into the overall stress. Even if (\ref{409}) can be validated as an
 extension of (\ref{gg} - \ref{dd})  to nonlinear regimes, we should
 remember that the second term in (\ref{409}) is much smaller than the
 first one (because we agreed that $\;E_{_{peak}}(\chi_{\textstyle{_2}})\,
 \ll\,E_{_{peak}}(\chi_{\textstyle{_1}})\;$). This results in two quandaries.
 The first one (not mentioned by Goldreich) is that a nonlinearity-caused
 non-smooth behaviour of $\;Q(\,\chi_{\textstyle{_1}}\,,\;E_{_{peak}}(
 \chi_{\textstyle{_1}})\,)\;$ will cause variations of the first term in
 (\ref{409}), which may exceed or be comparable to the entire second term.
 The second one (mentioned in the afore quoted passage from Goldreich) is
 the phenomenon of nonlinear superposition, i.e., the fact that the
 smaller-amplitude tidal harmonic has a higher dissipation rate (and,
 therefore, a larger phase lag) whenever the peak of this harmonic is
 reinforcing the peak of the principal mode. Under all these circumstances,
 fitting experimental data to (\ref{409}) will become a risky
 business. Specifically, it will become impossible to reliably measure the
 frequency dependence of the second quality factor; therefore the entire
 notion of the quality factor will, in regard to the second frequency, become
 badly defined.

 We shall not dwell on this topic in quantitative detail, leaving it for a future work.
 The only mentioning it is to draw the readers' attention to the existing difficulty
 stemming from the shortcomings of the extension of (\ref{gg} - \ref{dd})  to nonlinear
 regimes. In what follows, we shall consider linear deformations only.

 \section{Darwin (1879) and Kaula (1964)}

 The potential produced at point $\eRbold=(R\,,\,\lambda\,,\,\phi)\,$ by a
 mass $\,M^*$ located~at~$\,\erbold^{\;*}=(r^*,\,\lambda^*\,,\,\phi^*)\,$~is
 \ba
 W(\eRbold\,,\,\erbold^{\;*})\,=\,-\,\frac{G\;M^*}{r^{\,*}}
 \sum_{{\it{l}}=2}^{\infty}\left(\frac{R}{r^{\;*}}\right)^{
 \textstyle{^{\it{l}}}}\sum_{m=0}^{\it l}\frac{({\it l} - m)!
 }{({\it l} + m)!}(2-\delta_{0m})P_{{\it{l}}m}(\sin\phi)P_{{
 \it{l}}m}(\sin\phi^*)\;\cos m(\lambda-\lambda^*)~~.~~~~~~
 \label{1}
 \ea
 When a tide-raising secondary located at $\,\erbold^{\;*}\,$ distorts the shape of the
 primary, the potential generated by the primary at some exterior point $\,\erbold\,$
 gets changed. In the linear approximation, its variation is:
 \ba
 U(\erbold)\;=\;\sum_{{\it l}=2}^{\infty}\;k_{\it l}\;\left(\,\frac{R}{r}\,
 \right)^{{\it l}+1}\;W_{\it{l}}(\eRbold\,,\;\erbold^{\;*})~~~,~~~~~~~
 ~~~~~~~~~~~~~~~~
 \label{2}
 \ea
 $\,k_{\it{l}}\,$ being the {\emph{l}}{\small{th}} Love number, $R\,$ now being the mean
 equatorial
 radius of the primary, $\,\eRbold\,=
 \,(R\,,\,\phi\,,\,\lambda)\,$ being a surface point, $\,\erbold^{\;*}\,=\,(r^*\,,\,\phi^*
 \,,\,\lambda^*)\,$ being the coordinates of the tide-raising secondary, $\,\erbold\,=\,(r
 \,,\,\phi\,,\,\lambda)\,$ being an exterior point located above the surface point $\,
 \eRbold\,$ at a radius $\,r\,\geq\,R\,$, and the longitudes being reckoned
 from a fixed meridian on the primary.

 Substitution of (\ref{1}) into (\ref{2}) entails
 \ba
 U(\erbold)\;=\;\,-\,{G\;M^*}
 \sum_{{\it{l}}=2}^{\infty}k_{\it l}\;
 \frac{R^{
 \textstyle{^{2\it{l}+1}}}}{r^{
 \textstyle{^{\it{l}+1}}}{r^{\;*}}^{
 \textstyle{^{\it{l}+1}}}}\sum_{m=0}^{\it l}\frac{({\it l} - m)!
 }{({\it l} + m)!}(2-\delta_{0m})P_{{\it{l}}m}(\sin\phi)P_{{
 \it{l}}m}(\sin\phi^*)\;\cos m(\lambda-\lambda^*)~~~.~~~~~
 \label{4}
 \ea
 A different expression for the tidal potential was offered by Kaula (1961,
 1964), who developed a powerful technique that enabled him to switch from
 the spherical coordinates to the Kepler elements $\,(\,a^*,\,e^*,\,\inc^*,\,
 \Omega^*,\,\omega^*,\,{\cal M}^*\,)\,$ and $\,(\,a,\,e,\,\inc,\,\Omega,\,
 \omega,\,{\cal M}\,)\,$ of the secondaries located at $\,\erbold^{\;*}\,$ and
 $\,\erbold\,$. Application of this technique to (\ref{4}) results in
 \ba
 \nonumber
 U(\erbold)\;=\;-\;\sum_{{\it
  l}=2}^{\infty}\;k_{\it l}\;\left(\,\frac{R}{a}\,\right)^{\textstyle{^{{\it
  l}+1}}}\frac{G\,M^*}{a^*}\;\left(\,\frac{R}{a^*}\,\right)^{\textstyle{^{\it
  l}}}\sum_{m=0}^{\it l}\;\frac{({\it l} - m)!}{({\it l} + m)!}\;
  \left(\,2\;\right. ~~~~~~~~~~~~~~~~~~~~~~~~~~~~~~~~~~~~~~~~~~~~~~~~~~~~~\\
                                   \label{3}\\
                                   \nonumber\\
                                    \nonumber
 ~~~~~\left.-\,\delta_{0m}\,\right)\,\sum_{p=0}^{\it
  l}F_{{\it l}mp}(\inc^*)\sum_{q=-\infty}^{\infty}G_{{\it l}pq}
  (e^*) \sum_{h=0}^{\it l}F_{{\it
  l}mh}(\inc)\sum_{j=-\infty}^{\infty}
  G_{{\it l}hj}(e)\;\cos\left[
  \left(v_{{\it l}mpq}^*-m\theta^*\right)-
  \left(v_{{\it l}mhj}-m\theta\right) \right]
 ~~_{\textstyle{_{\textstyle ,}}}
 \ea
 where
 \ba
 v_{{\it l}mpq}^*\;\equiv\;({\it l}-2p)\omega^*\,+\,
 ({\it l}-2p+q){\cal M}^*\,+\,m\,\Omega^*~~~,
 \label{6}
 \ea
  \ba
 v_{{\it l}mhj}\;\equiv\;({\it l}-2h)\omega\,+\,
 ({\it l}-2h+j){\cal M}\,+\,m\,\Omega~~~,
 \label{7}
 \ea
 and $\theta\,=\,\theta^*\,$ is the sidereal angle.

 While (\ref{3}) and (\ref{4}) are equivalent for an idealised elastic planet with
 an instant response of the shape, the situation becomes more involved when
 dissipation-caused delays come into play. Kaula's expression (\ref{3}),
 as well as its truncated, Darwin's version,\footnote{~While the treatment by Kaula
 (1964) entails the infinite Fourier series (\ref{3}), the developments by Darwin
 (1879) furnish its partial sum with $\,|{\it{l}}|,\,|q|,\,|j|\,\leq\,2\,.$ For a
 simple introduction into Darwin's method see Ferraz-Mello, Rodr{\'{\i}}guez \&
 Hussmann (2008).} is capable of accommodating separate phase lags for each harmonic
 involved:
 \ba
 \nonumber
 U(\erbold)\;=\;-\;\sum_{{\it
 l}=2}^{\infty}\;k_{\it l}\;\left(\,\frac{R}{a}\,\right)^{\textstyle{^{{\it
 l}+1}}}\frac{G\,M^*}{a^*}\;\left(\,\frac{R}{a^*}\,\right)^{\textstyle{^{\it
 l}}}\sum_{m=0}^{\it l}\;\frac{({\it l} - m)!}{({\it l} + m)!}\;
 \left(\,2\;-\right. ~~~~~~~~~~~~~~~~~~~~~~~~~~~~~~~~~~~~~~~~~~~~~~~~~\\
                                   \label{5}\\
                                   \nonumber\\
                                    \nonumber
 \left.\delta_{0m}\,\right)\,\sum_{p=0}^{\it
 l}F_{{\it l}mp}(\inc^*)\sum_{q=-\infty}^{\infty}G_{{\it l}pq}
 (e^*)
 \sum_{h=0}^{\it l}F_{{\it
 l}mh}(\inc)\sum_{j=-\infty}^{\infty}
 G_{{\it l}hj}(e)\;\cos\left[
 \left(v_{{\it l}mpq}^*-m\theta^*\right)-
 \left(v_{{\it l}mhj}-m\theta\right)-
 \epsilon_{{\it l}mpq} \right]
 ~~_{\textstyle{_{\textstyle .}}}
  \ea
 where
 \ba
 \epsilon_{{\it l}mpq}=\left[\,({\it l}-2p)\,\dot{\omega}^*\,+\,
 ({\it l}-2p+q)\,\dot{\cal{M}}^*\,+\,m\,(\dot{\Omega}^*\,-\,\dot{\theta}^*)\,
 \right]\,\Delta t_{\it{l}mpq}=\,\omega^*_{\it{l}mpq}\,\Delta t_{\it{l}mpq}
 =\,\pm\,\chi^*_{\it{l}mpq}\,\Delta t_{\it{l}mpq}~~,~~~~
 \label{8}
 \ea
 is the phase lag interconnected with the quality factor via $\;Q_{{\it{l}}mpq}
 \,=\,\cot\,|\epsilon_{{\it{l}}mpq}|\;$. The tidal harmonic $\,\omega^*_{\it{l}mpq}
 \,$ introduced in (\ref{8}) is
 \ba
 \omega^*_{{\it l}mpq}\;\equiv\;({\it l}-2p)\;\dot{\omega}^*\,+\,({\it l}-
 2p+q)\;\dot{\cal{M}}^*\,+\,m\;(\dot{\Omega}^*\,-\,\dot{\theta}^*)\;~~,~~~
 \label{9}
 \ea
 while the positively-defined quantity
 \ba
 \chi^*_{{\it l}mpq}\,\equiv\,|\,\omega^*_{{\it l}mpq}\,|\,=\,|\,({\it
 l}-2p)\,\dot{\omega}^*\,+\,({\it l}-2p+q)\,\dot{\cal{M}}^*\,+\,m\,(\dot{\Omega}^*
 \,-\,\dot{\theta}^*)\;|~~~~~
 \label{10}
 \ea
 is the actual physical $\,{{\it l}mpq}\,$ frequency excited by the tide in the primary.
 The corresponding positively-defined time delay $\,\Delta t_{\it{l}mpq}\,$ depends on
 this physical frequency, the functional forms of this dependence being different for
 different materials.

 Formulae (\ref{3}) and (\ref{5}) constitute the principal result of Kaula's theory of
 tides. Most importantly, Kaula's formalism imposes no {\emph{a priori}} constraint on
 the form of frequency-dependence of the lags.

 \section{The Darwin-Kaula-Goldreich expansion\\
 for the tidal torque}

 Now we are prepared to calculate the planet-perturbing tidal torque. Since in what
 follows we shall dwell on the low-inclination case, it will be sufficient to derive
 the torque's component orthogonal to the planetary equator:
 \ba
 {\tau}\;=\;-\;{M}\;\frac{\partial U(\erbold)}{\partial\theta}\;\;\;,
 \label{29}
 \ea
 $M\,$ being the mass of the tide-disturbed satellite, and the ``minus" sign
 emerging due to our choice not of the astronomical but of the physical sign
 convention. Adoption of the latter convention implies the emergence of
 a ``minus" sign in the expression for the potential of a point mass:
 $\;-\,Gm/r\,$. This ``minus" sign then shows up on the right-hand sides of
 (\ref{1}), (\ref{4}), (\ref{3}), and (\ref{5}). It is then compensated by the ``minus" sign standing in
 (\ref{29}).

 The right way of calculating $\,{\partial U(\erbold)}/{\partial \theta}\,$
 is to take the derivative of (\ref{5}) with respect to $\,\theta\,$, and
 then\footnote{~Be mindful that our intention here is to differentiate not $\;\cos\left[
 \left(v_{{\it l}mpq}^*-m\theta^*\right)-
 \left(v_{{\it l}mhj}-m\theta\right)\right]\;$ but $\;\cos\left[
 (v_{{\it l}mpq}^{*{(delayed)}}-m{\theta^*}^{(delayed)})-
 \left(v_{{\it l}mhj}-m\theta\right)\right]\;$. Hence the said
 sequence of operations.} to get rid of the sidereal
 angle completely, by imposing the constraint $\,\theta^*\,=\,\theta\,$. This will
 yield:
 \ba
 \nonumber
 {\tau}= -\,\sum_{{\it
 l}=2}^{\infty}k_{\it l}\left(\frac{R}{a}\right)^{\textstyle{^{{\it
 l}+1}}}\frac{G\,M^*\,M}{a^*}\left(\frac{R}{a^*}\right)^{\textstyle{^{\it
 l}}}\sum_{m=0}^{\it l}\frac{({\it l} - m)!}{({\it l} + m)!}
 2m\;\sum_{p=0}^{\it
 l}F_{{\it l}mp}(\inc^*)\sum_{q=-\infty}^{\infty}G_{{\it l}pq}
 (e^*)~~~~~~~~~~~~~~~\\
                                   \nonumber\\
 \sum_{h=0}^{\it l}F_{{\it l}mh}(\inc)\sum_{j=-\infty}^{\infty} G_{{\it l}hj}(e)
 \;\sin\left[\,
 v^*_{{\it l}mpq}\,-\;v_{{\it l}mhj}\,-\;\epsilon_{{\it l}mpq}\,\right]
 ~~_{\textstyle{_{\textstyle ,}}}~~~~~~~~~~~~~~~~~~~~~~~~~~~~~~~~~~~
 \label{30}
 \ea
 In the case of the tide-raising satellite coinciding with the tide-perturbed
 one, all the elements become identical to their counterparts with an asterisk.
 For a primary body not in a tidal lock with its satellite,\footnote{~With $\alpha$
 denoting the librating angle, the locking condition reads: $\;\,\theta\,=\,\Omega\,+\,
 \omega\,+\,{\cal M}\,+\,180^o\,+\,\alpha\,+\,O(i^2)\;\,$. Insertion thereof into
 (\ref{9}) results in: $\;\omega^*_{{\it l}mpq}\;\equiv\;({\it l}-2p-m)\;\dot{\omega}^*
 \,+\,({\it l}-2p+q-m)\;\dot{\cal{M}}^*\,$,
 where we have neglected $\,\;-m\dot{\alpha}\,$ on account of $\,{\alpha}\,$ being
 extremely small. Clearly, the indices can assume more than one set of values
 corresponding to one tidal frequency. This way, the case of libration is more involved
 than that of tidal despinning, and deserves a separate consideration.} it is
 sufficient to limit our consideration to the constant part of the torque,\footnote{~The
 tide-raising and tidally-perturbed satellites being the same body does {\emph{not}} yet
 mean that the indices $\,(p\,,\,q)\,$ coincide with with $\,(h\,,\,j)\,$. These are two
 independent sets of indices, wherewith the terms of two Fourier series are numbered,
 expression (\ref{30}) being a product of those two series. This product
 contains a constant part, as well as short-period terms proportional to $\;\dot{\cal{M}}\;$
 and long-period terms proportional to $\,\dot{\omega}\,$. The short-period
 terms get averaged out over a period of the tidal flexure, while the
 long-period terms get averaged out over longer times, {\emph{provided the periapse is
 precessing and not librating}}. Expression (\ref{31}) furnishes the constant part of the torque.
 Fortunately, this is sufficient for our further calculations.} a
 part for which the indices $\,(p\,,\,q)\,$ coincide with
 $\,(h\,,\,j)\,$, and therefore $\,v_{{\it l}mhj}\,$ cancels with $\,v_{{\it l}mpq}^*\,$.
 This will give us:
 \ba
 {\tau}~=~\sum_{{\it{l}}=2}^{\infty}2~k_{\it l }~G~M^{\textstyle{^{2}}}~R^{\textstyle{^{2{\it{
 l}}\,+\,1}}}a^{\textstyle{^{-\,2\,{\it{l}}\,-\,2}}}\sum_{m=0}^{\it l}
 \frac{({\it{l}}\,-\,m)!}{({\it{l}}\,+\,m)!}
 \;m\;\sum^{\it l}_{p=0}
 \;F^{\textstyle{^{2}}}_{{\it{l}}mp}(\inc)\sum^{\it \infty}_{q=-\infty}G^{\textstyle{^{2}}}_{{\it{l}}pq}(e)
 \;\sin\epsilon_{{\it{l}}mpq}\;\;\;.~~~
 \label{31}
 \ea
 The expression gets considerably simplified if we restrict ourselves to
 the case of $\,{\it l}\,=\,2\,$. Since $\,0\,\leq\,m\,\leq\,{\it l}\,$,
 and since $\,m\,$ enters the expansion as a multiplier, we see that only
 $\,m\,=\,1\,,\,2\,$ actually matter. As $\,0\,\leq\,p\,\leq\,{\it l}\,$,
 we are left with only six relevant $\,F$'s, those corresponding to
 $\;(\it{l}mp)\,=\,$ (210), (211), (212), (220), (221), and (222). By a
 direct inspection of the table of $\,F_{\it{l}mp}\,$ we find that five
 of these six functions happen to be $\,O(\inc)\,$ or $\,O(\,\inc^2\,)\,$,
 the sixth one being $\,F_{220}\,=\,\frac{\textstyle 3}{\textstyle
 4}\,\left(\,1\,+\,\cos\inc\,\right)^2\,=\,3\,+\,O(\inc^2)\,$. Thus we
 obtain, in the leading order of $\,\inc\;$:
 \ba
 {\tau}_{\textstyle{_{\textstyle_{\textstyle{_{l=2}}}}}}~=~\frac{3}{2}
 ~\sum_{q=-\infty}^{\infty}~G~M^{\textstyle{^{2}}}~~R^{\textstyle{^{5}}}\;
 a^{-6}\;G^{\textstyle{^{2}}}_{\textstyle{_{\textstyle{_{20\mbox{\it{q}}}}}}}
 (e)\;k_{{2}}\;\sin\epsilon_{\textstyle{_{\textstyle{_{220\mbox{\it{q}}}}}}}
 \;+\;O(\inc^2/Q)\;\;\;.
 \label{32}
 \ea
 The leading term of the expansion is
 \ba
 {\tau}_{\textstyle{_{\textstyle{_{\textstyle{_{2200}}}}}}}~=~\frac{3}{2}
 ~G\,M^2~k_2~R^{\textstyle{^5}}\;a^{{{-\,6}}}
 \sin\epsilon_{\textstyle{_{2200}}}\;\;\;.
 \label{33}
 \ea
 Switching from the lags to quality factors via formula\footnote{~The phase lag
 $\,\epsilon_{\it{l}mpq}\,$ is introduced in (\ref{8}), while the
 tidal
 harmonic $\,\omega_{\it{l}mpq}\,$ is given by (\ref{9}). The quality factor
 $\,Q_{\it{l}mpq}\,=\,|\,\cot \epsilon_{\it{l}mpq}\,|$ is, for physical reasons,
 positively defined. Hence the
 multiplier $\,\mbox{sgn}\,\omega_{\it{l}mpq}\,$ in (\ref{34}). (As ever, the
 function $\,\mbox{sgn}(x)\,$ is defined to assume the values $\,+1\,$, $\,-1\,$,
 or $\,0\,$ for positive, negative, or vanishing $\,x\,$, correspondingly.)

 Mind that no factor of two appears in (\ref{Q} - \ref{epsilon}), because $\epsilon$ is a
 phase lag, not a geometric angle.}
 \ba
 Q_{\it{l}mpq}\,=\,|\,\cot \epsilon_{\it{l}mpq}\,|\;\;\;,
 \label{Q}
 \ea
 we obtain:
 \ba
 \sin\epsilon_{\textstyle{_{\textstyle{_{{{{\it{l}}mpq}}}}}}}=\,
 \sin|\epsilon_{\textstyle{_{\textstyle{_{{\it{l}}mpq}}}}}|\;\,\mbox{sgn}\,
 \omega_{\textstyle{_{\textstyle{_{{\it{l}}mpq}}}}}=\,\frac{\mbox{sgn}\,
 \omega_{\textstyle{_{\textstyle{_{{\it{l}}mpq}}}}}\;}{\sqrt{{\textstyle 1~+~\cot^2
 \epsilon_{\textstyle{_{\textstyle{_{{\it{l}}mpq}}}}}}}}=\;\frac{\mbox{sgn}\,
 \omega_{\textstyle{_{\textstyle{_{{\it{l}}mpq}}}}}\;}{\sqrt{{\textstyle 1~+~
 Q^{\textstyle{^{2}}}_{\textstyle{_{\textstyle{_{{\it{l}}mpq}}}}}}}}
 =~\frac{~\mbox{sgn}\,
 \omega_{\textstyle{_{\textstyle{_{{\it{l}}mpq}}}}}~}{Q_{
 \textstyle{_{\textstyle{_{{\it{l}}mpq}}}}}}+O(Q^{-3})~~,~~~
 \label{34}
 \label{epsilon}
 \ea
 whence
 \ba
 \nonumber
 {\tau}_{\textstyle{_{\textstyle_{\textstyle{_{l=2}}}}}}~=~
 \frac{3}{2}~\sum_{q=-\infty}^{\infty}~G~M^2~\;R^{\textstyle{^{5}}}\;a^{-6}
 \;G^{\textstyle{^{2}}}_{\textstyle{_{20\mbox{\it{q}}}}}(e)\;
 k_{\textstyle{_2}}\;\frac{~\mbox{sgn}\,
 \omega_{\textstyle{_{220\mbox{\it{q}}}}}\,}{
 Q_{\textstyle{_{\textstyle{_{220\mbox{\it{q}}}}}}}}\;\,
 \,+\,O(\inc^2/Q)\,+\,O(Q^{-3})\;\;\;.
 \label{}
 \ea
 Now, let us simplify the sign multiplier. If in expression (\ref{9}) for
 $\omega_{\textstyle{_{{\it{l}}mpq}}}$ we get rid of the
 redundant asterisks, replace\footnote{~While in the undisturbed two-body
 setting $\,{\cal{M}}\,=\,{\cal{M}}_0+n\,(t-t_0)\,$ and $\,\dot{\cal{M}}
 =n\,$, under perturbation these relations get altered. One possibility is to
 introduce (following Tisserand 1893) an {\emph{osculating mean motion}}
 $\,n(t)\,\equiv\,\sqrt{\mu/a(t)^3}\,$, and to stick to this definition under
 perturbation. Then the mean anomaly will evolve as
 $\,{\cal{M}}\,=\,{\cal{M}}_0(t)
 +\int_{t_o}n(t)\,dt\,$, whence $\,\dot{\cal{M}} =\dot{\cal{M}}_0(t)+n(t)\,$.

 Other possibilities include introducing an {\emph{apparent}} mean motion, i.e.,
 defining $\,n\,$ either as the mean-anomaly rate $\,d{\cal M}/dt\,$, or as the
 mean-longitude rate $\,dL/dt\,=\,d\Omega/dt\,+\,d\omega/dt\,+\,d{\cal{M}}/dt\,$
 (as was done by Williams et al. 2001). While the first-order perturbations in $a(t)$ and in the osculating mean motion
 $\sqrt{\mu/a(t)^3}$ do not have secular rates, the epoch terms typically do have
 secular rates. Hence the difference between the apparent mean motion defined as
 $dL/dt$ (or as $d{\cal M}/dt$) and the osculating mean motion $\sqrt{\mu/a(t)^3}$.
 I am thankful to
 James G. Williams for drawing my attention to this circumstance (J.G. Williams,
 private communication).
 } $\;\dot{\cal{M}}$ with $\dot{\cal{M}}_0+n\approx n$, and set ${\it{l}}=m=2$
 and $p=0$, the outcome will be:
 \ba
 \nonumber
 \mbox{sgn}\,\omega_{\textstyle{_{220\mbox{\it{q}}}}}\;=\;
 \mbox{sgn}\,\left[\,2\;\dot{\omega}\,+\,(2+q)\;n\,+\,2\,\dot{\Omega}-\,2\,\dot{\theta}
 \,\right]\;=\;\mbox{sgn}\,\left[\,\dot{\omega}\,+\,\left(1\,+\,
 \frac{\textstyle q}{\textstyle 2}\,\right)\;n\,+\,\dot{\Omega}-\,\dot{\theta}
 \,\right]~~~.
 \label{}
 \ea
 As the node and periapse precessions are slow,
 the above expression may be simplified to
 \ba
 \nonumber
 \mbox{sgn}\,\left[\,\left(1\,+\,
 \frac{\textstyle q}{\textstyle 2}\,\right)\;n\,-\,\dot{\theta}
 \,\right]~~~.
 \label{}
 \ea
 All in all, the approximation for the torque assumes the form:
  \ba
 {\tau}_{\textstyle{_{\textstyle_{\textstyle{_{l=2}}}}}}\,=\,
 \frac{3}{2}~\sum_{q=-\infty}^{\infty}~G~M^2~\;R^{\textstyle{^{5}}}\;a^{-6}
 \,G^{\textstyle{^{2}}}_{\textstyle{_{20\mbox{\it{q}}}}}(e)\;
 k_{\textstyle{_2}}\;{Q^{\textstyle{^{-1}}}_{\textstyle{_{\textstyle{_{220\mbox{\it{q}}}}}}}}
 \;~\mbox{sgn}\,\left[\,\left(1\,+\,
 \frac{\textstyle q}{\textstyle 2}\,\right)\;n\,-\,\dot{\theta}
 \,\right]
 +O(\inc^2/Q)+O(Q^{-3})\;\;.~~~
 \label{35}
 \label{exp}
 \ea
 That the sign of the right-hand side in the above formula is correct can be
 checked through the following obvious observation: for a sufficiently high
 spin rate $\,\dot{\theta}\,$ of the planet, the multiplier $\,\mbox{sgn}\,
 \left[\,\left(1\,+\,\frac{\textstyle q}{\textstyle 2}\,\right)\;n\,-\,
 \dot{\theta}\,\right]\,$ becomes negative. Thereby the overall expression
 for $\,{\tau}_{\textstyle{_{\textstyle_{\textstyle{_{l=2}}}}}}\,$
 acquires a ``minus" sign, so that the torque points out in the direction
 of rotation opposite to the direction of increase of the sidereal
 angle $\,\theta\,$. This is exactly how it should be, because for a
 fixed $\,q\,$ and a sufficiently fast spin the $\,q$'s component of the
 tidal torque must be decelerating and driving the planet to synchronous
 rotation.

 Expansion (\ref{exp}) was written down for the first time, without proof, by
 Goldreich \& Peale (1966). A schematic proof was later offered by Dobrovolskis (2007).
 %

 \section{Can the quality factor scale as a positive power of
the tidal frequency?}

 As of now, the functional form of the dependence $\,Q(\chi)\,$ for Jovian
 planets remains unknown. For terrestrial planets, the model $\,Q\,\sim\,1/\chi\,
 $ is definitely incompatible with the geophysical data. A convincing volume of
 measurements firmly witnesses that $\,Q\,$ of the mantle scales as the tidal
 frequency to a {\emph{positive}} fractional power:
 \ba
 Q
 \;=\;{\cal E}^{\alpha}\;\chi^{\alpha}
 \;\;\;,
 ~~~~~\mbox{where}\;\;\;\alpha\,=\,0.3\,\pm\,0.1\;\;\;,
 \label{55}
 \ea
 ${\cal E}\,$ being an integral rheological parameter with dimensions of time. This rheology
 is incompatible with the postulate of frequency-independent time-delay. Therefore an
 honest calculation should be based on averaging the Darwin-Kaula-Goldreich formula
 (\ref{35}), with the actual scaling law (\ref{55}) inserted therein, and with the
 appropriate dependence $\,\Delta t_{\textstyle{_{lmpq}}}(\,\chi_{\textstyle{_{lmpq}}}\,)\,$
 taken into account.\footnote{~For the dependence of $\,\Delta t_{\textstyle{_{lmpq}}}\,$
 upon $\,\chi_{\textstyle{_{lmpq}}}\,$ see Efroimsky \& Lainey 2007.}

  \subsection{The ``paradox"}

 Although among geophysicists the scaling law (\ref{55}) has long become common knowledge,
 in the astronomical community it is often met with prejudice. The prejudice stems from
 the fact that, in the expression for the torque, $\,Q\,$ stands in the denominator:
 \ba
 \tau\;\sim\;\frac{1}{Q}\;\;\;\,.\,~~~~~~~~~~~~~~~~~
 \label{torque}
 \ea
 At the instant of crossing the synchronous orbit, the principal tidal frequency
 $\,\chi_{\textstyle{_{\textstyle{_{2200}}}}}\,$ becomes nil, for which reason
 insertion of
 \ba
 Q
 \;\sim\;\chi^{\alpha}
 \;\;\;,
 ~~~~~\alpha\,>\,0\;\;\;
 \label{}
 \ea
 into (\ref{torque}) seems to entail an infinitely large torque at the
 instant of crossing:
 \ba
 \tau\;\sim\;\frac{1}{Q}\;\sim\;\frac{1}{\chi^{\alpha}}\;\rightarrow\;
 \infty
 ~~,~~~~~{\mbox{for}}~~~\chi\;\rightarrow\;0~~~,
  \label{infinity}
 \ea
 a clearly unphysical result.

 Another, very similar objection to (\ref{55}) originates from the fact that
 the quality factor is inversely proportional to the phase shift: $\,Q\,\sim\,
 1/\epsilon\,$. As the shift (\ref{8}) vanishes on crossing the synchronous orbit,
 one may think that the value of the quality factor must, effectively, approach
 infinity. On the other hand, the principal tidal frequency vanishes on crossing
 the synchronous orbit, for which reason (\ref{55}) makes the quality factor
 vanish. Thus we come to a contradiction.

 For these reasons, the long-entrenched opinion is that these models
 introduce discontinuities into the expression for the torque, and can thus be considered as
 unrealistic.

 It is indeed true that, while law (\ref{55}) works over scales shorter than
 the Maxwell time (about $\,10^2$ yr for most minerals), it remains subject
 to discussion in regard to longer timescales. Nonetheless, it should be
 clearly emphasised that the infinities emerging at the synchronous-orbit
 crossing can in no way disprove any kind of rheological model. They can
 only disprove the
 flawed mathematics whence they provene.

 \subsection{A case for reasonable doubt}

 To evaluate the physical merit of the alleged infinite-torque ``paradox",
 recall the definition of the quality factor. As part and parcel of the
 linearity approximation, the overall damping inside a body is expanded
 in a sum of attenuation rates corresponding to each periodic disturbance:
 \ba
 \langle\,\dot{E}\;\rangle\;=\;\sum_{i}\;\langle\,
 \dot{E}(\chi_{\textstyle{_i}})\;\rangle
 \label{407}
 \ea
 where, at each frequency $\,\chi_i\,$,
 \ba
 \langle\,\dot{E}(\chi_{\textstyle{_i}})~\rangle~=~-~2~\chi_{\textstyle{_i}}~
 \frac{\,\langle\,E(\chi_{\textstyle{_i}})~\rangle\,}{Q(\chi_{\textstyle{_i}})
 }~=\,\;-\;\chi_{\textstyle{_i}}\;\frac{\,E_{_{peak}}(\chi_{\textstyle{_i}})
 \,}{Q(\chi_{\textstyle{_i}})}\;\;\;,
 \label{408}
 \ea
 $\langle\,.\,.\,.\,\rangle~$ designating an average over a flexure cycle,
 $\,E(\chi_{\textstyle{_i}})\,$ denoting the energy of deformation at
 the frequency $\,\chi_{\textstyle{_i}}\,$, and $Q(\chi_{\textstyle{_i}})\,$
 being the quality factor of the medium at this frequency.

 This definition by itself leaves enough room for doubt in the above ``paradox".
 As can be seen from (\ref{408}), the dissipation rate is proportional not to
 $\;1/Q(\chi)\;$ but to $\;\chi/Q(\chi)\;$. This way, for the dependence $\,Q
 \,\sim\,\chi^{\alpha}\,$, the dissipation rate $\,\langle\dot{E}\rangle\,$
 will behave as $\,\chi^{1-\alpha}\,\;$. In the limit of $\,\chi\,\rightarrow
 \,0\,$, this scaling law portends no visible difficulties, at least for the
 values of $\,\alpha\,$ up to unity. While raising $\,\alpha\,$ above unity may
 indeed be problematic, there seem to be no fundamental obstacle to having
 materials with positive $\,\alpha\,$ taking values up to unity. So far, such
 values of $\,\alpha\,$ have caused no paradoxes, and there seems to be no reason
 for any infinities to show up.


 \subsection{The phase shift and the quality factor}

 As another preparatory step, we recall that, rigorously speaking, the
 torque is proportional not to the phase shift $\,\epsilon\,$ itself but
 to $\,\sin\epsilon\,$. From (\ref{34}) and (\ref{55}) we obtain:
 \ba
 |\,\sin\epsilon\,|\;=\;\frac{
 1}{\textstyle\sqrt{1\,+\,Q^{\textstyle{^2}}}}\;=\;\frac{1}{\sqrt{1\,+\;{\cal E}^{
 \textstyle{^{2
 \alpha}}}\;{\chi}^{\textstyle{^{2\alpha}}}\;}}\;\;\;.
 \label{shift}
 \ea
 We see that only for large values of $\,Q\,$ one can approximate $\;\,|\,
 \sin\epsilon\,|\;\,$ with $\,1/Q\,$ (crossing of the synchronous orbit
 {\emph{not}} being the case). Generally, in any expression for the
 torque, the factor $\,1/Q\,$ must always be replaced with $\,1/\sqrt{1\,+\,
 Q^2}\,\;$. Thus instead of (\ref{torque}) we must write:
 \ba
 \tau\;\sim\;|\,\sin\epsilon\,| \;=\;\frac{1}{\sqrt{1\,+\,Q^{\textstyle{^2}}\;}}\;=\;\frac{1}{\sqrt{1\,+\;{\cal E}^{
 \textstyle{^{2
 \alpha}}}\;{\chi}^{\textstyle{^{2\alpha}}}\;}}\;\;\;\;\,,\,~~~~~~~~~~~~~~~~~
 \label{relation}
 \ea
 ${\cal E}\,$ being a dimensional constant from (\ref{55}).

 Though this immediately spares us from the fake infinities at $\,\chi\,
 \rightarrow\,0\,$, we still are facing this strange situation: it follows
 from (\ref{shift}) that, for a positive $\,\alpha\,$ and vanishing
 $\,\chi\,$, the phase lag $\,\epsilon\,$ must be approaching $\,\pi/2\,$,
 thereby inflating the torque to its maximal value (while on physical
 grounds the torque should vanish for zero $\,\chi\,$). Evidently, some
 important details are still missing from the picture.

  \subsection{The stone rejected by the builders}


 To find the missing link, recall that Kaula (1964) described tidal damping
 by employing the method suggested by Darwin (1880): he accounted for
 attenuation by merely adding a phase shift to every harmonic involved --
 an empirical approach intended to make up for the lack of a consistent
 hydrodynamical treatment with viscosity included. It should be said,
 however, that prior to the work of 1880 Darwin had published a less known
 article (Darwin 1879), in which he attempted to construct a
 self-consistent theory, one based on the viscosity factor of the mantle,
 and not on empirical phase shifts inserted by hand. Darwin's conclusions
 of 1879 were summarised and explained in a more general mathematical
 setting by Alexander (1973).

 The pivotal result of the self-consistent hydrodynamical study is the
 following. When a variation of the potential of a tidally disturbed planet,
 $\,U(\erbold)\,$, is expanded over the Legendre functions $\,P_{{\it{l}}m}(
 \sin\phi)\,$, each term of this expansion will acquire not only a phase lag
 but also a factor describing a change in amplitude. This forgotten factor,
 derived by Darwin (1879), is nothing else but $\;\,\cos\epsilon\;$. Its
 emergence should in no way be surprising if we recall that the damped,
 forced harmonic oscillator
 \ba
 \ddot{x}\;+\;2\;\gamma\;\dot{x}\;+\;\omega^2_o\,x\;=\;F\;e^{\inc\,\lambda\,t}
 \label{}
 \ea
 evolves as
 \ba
 x(t)~=~C_1~\,e^{\textstyle{^{(\,-\,\gamma\,+\,\inc\,\sqrt{\omega_o^2-\lambda^2
 \,}\,)\;t}}}\,+\;
 C_2~\,e^{\textstyle{^{(\,-\,\gamma\,+\,\inc\,\sqrt{\omega_o^2-\lambda^2\,}\,)
 ~t}}}\,+~\frac{F~\cos\epsilon}{\omega_o^2\,-\,\lambda^2}~\,e^{\textstyle{^{
 \inc\,(\lambda\,t\,-\,\epsilon)}}}\;\;\;,
 \label{solution}
 \ea
 where the phase lag is
 \ba
 \tan\epsilon\;=\;2\;\gamma\;\lambda\;\left(\,\omega_o^2\,-\;\lambda^2\,\right)
 \;\;\;,
 \label{}
 \ea
 and the first two terms in (\ref{solution}) are damped away in
 time.\footnote{~As demonstrated by Alexander (1973), this example indeed has relevance to
 the hydrodynamical theory of Darwin, and is not a mere illustration. Alexander (1973)
 also explained that the emergence of the $\,\cos\epsilon\,$ factor is generic. (Darwin
 (1879) had obtained it in the simple case of $\,{\it l}\,=\,2\,$ and for a special value
 of the Love number: $\,k{\it{_{l}}}=\,1.5\,$.)

 A further investigation of this issue was undertaken in a comprehensive work
 by Churkin (1998), which unfortunately has never been published in English
 because of a tragic death of its Author. In this preprint, Churkin explored the
 frequency-dependence of both the Love number $\,k_2\,$ and the quality factor
 within a broad variety of rheological models, including those of Maxwell and
 Voight. It follows from Churkin's formulae that within the Voight model the
 dynamical $\,k_2\,$ relates to the static one as $\,\cos \epsilon\,$. In the
 Maxwell and other models, the ratio approaches $\,\cos \epsilon\,$ in the
 low-frequency limit.}

 In the works by Darwin's successors, the allegedly irrelevant factor of $\,\cos
 \epsilon\,$ fell through the cracks, because the lag was always asserted to be
 small. In reality, though, each term in the Fourier expansions (\ref{5}),
 (\ref{30} - \ref{33}), and (\ref{35}) should be amended with $\,\cos\epsilon_{
 \textstyle{_{\textstyle{_{{\it{l}}mpq}}}}}\,$. For the same reason, instead of
 (\ref{relation}), we should write down:
 \ba
 \tau\;\sim\;|\,\cos\epsilon\;\,\sin\epsilon\,|\;=\;\frac{Q}{\sqrt{1\,+\,
 Q^{\textstyle{^2}}\;}}\;\frac{1}{\sqrt{1\,+\,Q^{\textstyle{^2}}\;}}\;=\;\frac{{\cal E}^{\textstyle{^{
 \alpha}}}\;{\chi}^{\textstyle{^{\alpha}}}}{1\,+\;{\cal E}^{\textstyle{^{2
 \alpha}}}\;{\chi}^{\textstyle{^{2\alpha}}}}\;\;\;\;\,,\,~
 \label{71}
 \ea
 Introducing of the extra multiplier $\,\cos\epsilon_{\textstyle{_{\textstyle{_{{\it{l}}mpq}}}}}\,$ is in fact an attempt to
 empirically endow the Love number with frequency dependence. A rigorous method of doing this will be presented elsewhere.
 Meanwhile, we would mention that amending of the Love number with the $\,\cos\epsilon_{\textstyle{_{\textstyle{_{{\it{l}}mpq}}}}}\,$ 
 factor works well for the Kelvin-Voigt body. For the Maxwell rheology, this method turns out to be correct only at low frequencies
 (lower than the inverse Maxwell time). 
 
 At this point, it would be tempting to conclude that, since (\ref{71}) vanishes in the limit of
 $\chi\rightarrow0\;$, {\emph{for any sign}} {\emph{of}} $\alpha\;$, then no paradoxes happens on the satellite's crossing the
 synchronous orbit. Sadly, this straightforward logic would be too simplistic.

 In fact, prior to saying that $\,\cos\epsilon\,\sin\epsilon\rightarrow0$, we must take into consideration one more
 subtlety missed so far. As demonstrated in the Appendix, taking the limit of $Q\rightarrow 0$ is a
 nontrivial procedure, because at small values of $\,Q\,$ the interconnection between the lag
 and the Q factor becomes very different from the conventional $Q=\cot|\epsilon|$. A laborious
 calculation shows that, for $\;Q<1-\pi/4\,$, the relation becomes:
  \ba
  \nonumber
  \sin\epsilon\,\cos\epsilon\,=\,\pm\;(3Q)^{1/3}\,\left[1-\frac{4}{5}(3Q)^{2/3}+O(Q^{4/3})\right]\;\;\;,
  \label{}
  \ea
 which indeed vanishes for $Q\rightarrow 0$. Both $\,\epsilon_{\textstyle{_{2200}}}\,$ and the
 appropriate component of the torque change their sign on the satellite crossing the synchronous orbit.

 So the main conclusion remains in force: nothing wrong happens on crossing the synchronous orbit,
 ~Q.E.D.\\

 \section{Conclusions}

 In the article thus far we have punctiliously spelled out some assumptions
 that often remain implicit, and brought to light those steps in calculations,
 which are often omitted as ``self-evident". This has helped us to explain that no
 ``paradoxes" ensue from the frequency-dependence $\;Q\,\sim\,\chi^{\alpha}\;\;,\;\;\alpha\,=\,
 0.3\,\pm\,0.1\;$, which is in fact the actual dependence found for the mantle and crust.

 This preprint is a pilot paper. A more comprehensive treatise on tidal torques
 is to be published. (Efroimsky \& Williams 2009)\\
~\\
~\\


 {\underline{\textbf{\Large{Acknowledgments}}}}\\
 ~\\
 I deeply thank Bruce Bills, Alessandra Celletti, Tony Dobrovolskis, Peter Goldreich,
 Shun-ichiro Karato, Valery Lainey, William Newman, Stan Peale, S. Fred Singer, and Gabriel Tobie -- the
 colleagues with whom I on many occasions had stimulating conversations on the theory of
 tides, and from whom I have learned much. My special gratitude goes to Sylvio Ferraz Mello
 and James G. Williams, who kindly offered numerous valuable comments on my
 text.\\

 ~\\


 \noindent
  {\underline{\textbf{\Large{Appendix.}}}}\\
  ~\\
  {\textbf{\large{The lag~and~the quality$\,$factor:~is~the~formula~$\boldmath{Q=\cot |\epsilonbold |}$~universal?}}}\\
  ~\\

  The interrelation between the quality factor $\,Q\,$ and the phase lag $\,\epsilon\,$ is
 long-known to be
 \ba
 Q\,=\,\cot |\epsilon |\;\;\;,
 \label{formula}
 \ea
 and its derivation can be found in many books. In Appendix A2 of Efroimsky \& Lainey(2007),
 that derivation is reproduced, with several details that are normally omitted in the literature. Among other things, we pointed out that the interrelation has exactly the form (\ref{formula})
 only in the limit of small lags. For large phase lags, the form of this relation will
 change considerably.

 Since in section 9 of the current paper we address the case of large lags, it would be worth reconsidering the derivation presented in Efroimsky \& Lainey (2007), and correcting a subtle omission made there. Before writing formulae, let us recall that, at each frequency $\,\chi\,$ in the spectrum of the deformation, the quality factor (divided by $\,2\,\pi\,$) is defined as the peak energy stored in the system divided by the energy damped over a cycle of flexure:
 \ba
 {Q}(\chi)\;\equiv\;-\;\frac{2\;\pi\;E_{peak}(\chi)}{\Delta E_{cycle}(\chi)}\;\;\;,
 \label{}
 \ea
 where $\,\Delta E_{cycle}(\chi)\,<\,0\,$ as we are talking about energy
 losses.\footnote{~We are considering flexure in the linear approximation. Thus at each
  frequency $\,\chi\,$ the appropriate energy loss over a cycle, $\,\Delta E_{cycle}(\chi)\,$, depends solely on the maximal energy stored at that same frequency, $\,E_{peak}(\chi)\,$.}

 An attempt to
 consider large lags (all the way up to $\,|\epsilon |\,=\,\pi/2\,$) sets the values of $\,Q/2\pi\,$
 below unity. As the dissipated energy cannot exceed the energy stored in a free oscillator, the
 question becomes whether the values of $\,Q/2\pi\,$ can be that small. To understand
 that they can, recall that in this situation we are considering an oscillator, which is not
 free but is driven (and is overdamped). The quality factor being much less than unity simply
 implies that the eigenfrequencies get damped away during less than one oscillation. Nonetheless,
 motion goes on due to the driving force.

 Now let us switch to the specific context of tides. The power $\,P\,$ exerted by a tide-raising secondary on its primary can be written 
 as
 \ba
 P\;=\;-\;\int \,\rho\;\Vbold\;\cdot\;\nabla W\;d^3x
 \label{A99999}
 \ea
 $\rho\,,\;\Vbold\,$, and $\,W\,$ signifying the density, velocity, and tidal potential in the small volume $~d^3x~$ of the primary.
 The mass-conservation law $~\nabla\cdot(\rho\Vbold)\,+\frac{\textstyle \partial \rho}{\textstyle\partial t}\,=\,0\,~$ enables one to
 shape the dot-product into the form of
 \ba
 \rho\,\Vbold\cdot\nabla W\,=\,
 \nabla\cdot(\rho\,
 \Vbold\,W)\,-\,\rho\,W\,\nabla\cdot\Vbold\,-\,\Vbold\,W\,\nabla\rho\;\;~.\;\;\;\;
 \label{}
 \ea
 Under the realistic assumption of the primary's incompressibility, the term with $\,\nabla\cdot\Vbold\,$ may be omitted. To get rid
 of the term with $\,\nabla \rho\,$, one has to accept a much stronger approximation of the primary being homogeneous.
 Then the power will be rendered by
 \ba
 P\;=\;-\;\int\,\nabla\,\cdot\,(\rho\;\Vbold\;W)\,d^3x
 \;=\;-\;\int\,\rho\;W\;\Vbold\,\cdot\,{\vec{\bf{n}}}\;\,dS\;\;\;,
 \label{A9}
 \ea
 ${\vec{\bf{n}}}\,$ being the outward normal and $\,dS\,$ being an element of the surface area of the primary. This expression for the
 power (pioneered, probably, by Goldreich 1963) enables one to calculate the work through radial displacements only, in neglect of
 horizontal motion. 
 
 We can write the power per unit mass, $~{\cal P}\equiv P/M~$, as:
 \ba
 {\cal P}\;=\;\left(-\,\frac{\partial W}{\partial r}\right)\;\Vbold\cdot{\vec{\bf{n}}}
 \;=\;\left(-\,\frac{\partial W}{\partial r}\right)\frac{d\zeta}{dt}\;\;\;,
 \label{}
 \ea
 $\zeta\,$ standing for the vertical displacement (which is, of course, delayed in time, compared to $\,W\,$).
 
 This power per unit mass is a good approximation for the energy dissipation rate in the body, $\,\dot{E}\,$. A tiny difference between 
 these quantities exists, because a part of the power is spent for decelerating or accelerating the spin. Neglecting this tiny 
 difference, we can write the above formula as 
 \ba
 \dot{E}\;=\;\left(-\,\frac{\partial W}{\partial r}\right)\;\Vbold\cdot{\vec{\bf{n}}}
 \;=\;\left(-\,\frac{\partial W}{\partial r}\right)\frac{d\zeta}{dt}\;\;\;,
 \label{}
 \ea
 
 The amount of energy dissipated over a time interval $\,(t_o\,,\;t)\,$ is then
 \ba
 \Delta{E}\;=\;\int^{t}_{t_o}\;\left(-\,\frac{\partial W}{\partial
 r}\right)\;d\zeta\;\;\;.
 \label{dissipation}
 \ea

 We shall consider the simple
 case of an equatorial moon on a circular orbit. At each point of the planet, the
 variable part of the tidal potential produced by this moon will read
  \ba
  W\;=\;W_o\;\cos \chi t\;\;\;,
  \label{A3}
  \label{469}
  \ea
 the tidal frequency being given by
 \ba
 \chi\,=\,2~|n\;-\;\omega_p|~~~.~~~
 \label{A3}
 \label{470}
 \ea
 Let $\,\mbox{g}\,$ denote the surface free-fall acceleration. An element of the planet's surface lying beneath the satellite's trajectory will then experience a vertical elevation  of
 \ba
 \zeta\;=\;h_2\;\frac{W_o}{\mbox{g}}\;\cos (\chi t\;-\;|\epsilon |)\;\;\;,
 \label{A4}
 \label{471}
 \ea
 $\,h_2\,$ being the corresponding Love number, and $\,|\epsilon |\,$ being the
 {\emph{positive}}\footnote{~Were we not considering the simple case of a circular orbit, then, rigorously
 speaking, the expression for $\,W\,$ would read not as $\,W_o\,\cos \chi t\,$ but
 as $\,W_o\,\cos \omega_{\textstyle{_{tidal}}} t\,$, the tidal frequency $\,\omega_{\textstyle{_{tidal}}}\,$ taking both positive and negative
 values, and the physical frequency of flexure being $\,\chi\,\equiv\,|\omega_{\textstyle{_{tidal}}}|\,$.
 Accordingly, the expression for $\,\zeta\,$ would contain not $\,\cos (\chi t\,-\,|\epsilon |)\,$ but
 $\,\cos (\omega_{\textstyle{_{tidal}}} t\,-\,\epsilon)\,$. As we saw in equation (24), the sign of $\,\epsilon\,$
 is always the same as that of $\,\omega_{\textstyle{_{tidal}}}\,$. For this reason, one may simply deal with
 the physical frequency $\,\chi\,\equiv\,|\omega_{\textstyle{_{tidal}}}|\,$ and with the absolute value of the phase lag, $\;| \epsilon |\;$.}
 phase lag, which for the principal tidal frequency is simply the double
 geometric angle $\,\delta\,$ subtended at the primary's centre between the
 directions to the secondary and to the main bulge:
 \ba
 |\epsilon|\;=\;2\;\delta\;\;\;.
 \label{dot}
 \ea
 Accordingly, the vertical velocity of this element of the planet's surface will
 amount to
  \ba
 u\;=\;\dot{\zeta}\;=\;-\;h_2\;\chi\;\frac{W_o}{\mbox{g}}\;\sin (\chi t
 \;-\;|\epsilon|)\;=\;-\;h_2\;\chi\;\frac{W_o}{\mbox{g}}\;\left(\sin \chi
 t\;\cos |\epsilon|\;-\;\cos \chi t\; \sin |\epsilon|\right)\;\;.\;\;
 \label{A5}
 \label{472}
 \ea
 The expression for the velocity has such a simple form because in this case the
 instantaneous frequency $\,\chi\,$ is constant. The satellite generates two bulges
 -- on the facing and opposite sides of the planet -- so each point of the surface
 is uplifted twice through a cycle. This entails the factor of two in the
 expression (\ref{470}) for the frequency. The phase in (\ref{dot}), too, is
 doubled, though the necessity of this is less evident, -- see footnote 4 in
 Appendix A1 to Efroimsky \& Lainey (2007).

 The energy dissipated over a time cycle $\,T\,=\,2\pi/\chi\,$, per
 unit mass, will, in neglect of horizontal displacements, be
 \ba
 \nonumber
 \Delta E_{_{cycle}} &=& \int^{T}_{0}u\left(-\,\frac{\partial W}{
 \partial r}\right)dt=
 \,-\left(-\,h_2\;\chi \frac{W_o}{\mbox{g}}\right)\,\frac{\partial W_o}{
 \partial r}\int^{t=T}_{t=0}\cos \chi t\,\left(\sin \chi t\,
 \cos |\epsilon|\,-\,\cos \chi t\, \sin |\epsilon|\right)dt\\
 \nonumber\\
 \nonumber\\
 &=&\,-\;h_2\;\chi\;\frac{W_o}{\mbox{g}}\;\frac{\partial W_o}{\partial r}\;\sin|\epsilon|\,
 \;\frac{1}{\chi}\;\int^{\chi t\,=\,2\pi}_{\chi t\,=\,0}\;\cos^2 \chi t\;\;d(\chi
 t)\;=\;-\;h_2\;\frac{W_o}{\mbox{g}}\;\frac{\partial W_o}{\partial r}\;\pi\;\sin|\epsilon|
 \;\;,\;\;\;~~~~~~~~~~~~~~~~~~~~
 \label{A6}
 \label{}
 \ea
 while the peak work carried out on the system during the cycle will read:
 \ba
 \nonumber
 E_{_{peak}}&=&\int^{T/4}_{|\epsilon|/\chi} u \left(-\,\frac{\partial W}{
 \partial r}\right)dt =
 \,-\left(-\,h_2\;\chi\,\frac{W_o}{\mbox{g}}\right)\frac{\partial W_o
 }{\partial r}\int^{t=T/4}_{t=|\epsilon|/\chi}\cos \chi t\,\left(\sin
 \chi t\,\cos |\epsilon|\,-\,\cos \chi t\,\sin |\epsilon|\right)dt\\
 \nonumber\\
 \nonumber\\
 &=&\;\chi\;h_2\;\frac{W_o}{\mbox{g}}\;\frac{\partial W_o}{\partial r}\;\left[\;
 \frac{\cos |\epsilon|}{\chi}\;\int^{\chi t\,=\,\pi/2}_{\chi t\,=\,|\epsilon|}
 \;\cos \chi t\;\sin \chi t\;\;d(\chi t)\;-\;\frac{\sin |\epsilon|
 }{\chi}\;\int^{\chi t\,=\,\pi/2}_{\chi t\,=\,|\epsilon|}\;\cos^2 \chi t
 \;\;d(\chi t)\;\right]\;\;.~~~~~~~~\,
 \ea
 In the appropriate expression in Appendix A1 to Efroimsky \& Lainey (2007), the lower limit of integration was erroneously set to be zero. To understand that in reality integration over $\,\chi\,t\,$ should begin from $\,|\epsilon|\,$, one should superimpose the plots of the two functions involved, $\,\cos \chi t\,$ and $\,\sin(\chi t\,-\,|\epsilon|)\,$. The maximal energy gets stored in the system after integration through the entire interval over which both functions have the same sign. Hence $\,\chi t=|\epsilon|\,$ as the lower limit.

 Evaluation of the integrals entails:
 \ba
 E_{peak}\;=\;h_2\;\frac{W_o}{\mbox{g}}\;\frac{\partial W_o}{\partial r}\;\left[\;\frac{1}{2}
 \;\cos |\epsilon|\;-\;\frac{1}{2}\;\left(\;\frac{\pi}{2}\;-\;|\epsilon|\;\right)\;\sin |\epsilon|\;\right]~~~~~~~~~~~~~~~~~~~
 ~~~~~~~~~~~~~~~~~~~~~~~~~~~~~~~~~~~~~~~~~~~~~~~~~~~~~~~
 \label{A7}
 \label{}
 \ea
 whence
 \ba
 Q^{-1}\;=\;\frac{-\;\Delta E_{_{cycle}}}{2\,\pi\,E_{_{peak}}}\;=\;\frac{1}{2\,\pi}
 \;\,\frac{\pi\;\sin |\epsilon|}{~\frac{\textstyle 1}{\textstyle 2}\;\cos |\epsilon|\;-\;
 \frac{\textstyle 1}{\textstyle 2}\;\left(\;\frac{\textstyle \pi}{\textstyle 2}\;-\;
 |\epsilon|\;\right)\;\sin |\epsilon|}\;=\;\frac{\tan |\epsilon|}{1\;-\;\left(\;\frac{\textstyle \pi}{\textstyle 2}\;-\;
 |\epsilon|\;\right)  \;\tan|\epsilon|}\;\;\;.~~~~~
 \label{A8}
 \label{}
 \ea
 As can be seen from (\ref{A8}), both the product $\,\sin\epsilon\,\cos\epsilon\,$ and
 the appropriate component of the torque attain their maxima when
 $\,Q\,=\,1\,-\,\pi/4\,$.

 Usually, $\,|\epsilon|\,$ is small, and we arrive at the customary expression
 \ba
 Q^{-1}\,=\,\tan|\epsilon|\;+\;O(\epsilon^2)\;\;\;.
 \label{customary}
 \ea
 In the opposite situation, when $\,Q\rightarrow 0$ and $\,|\epsilon| \rightarrow \pi/2\,$,
 it is convenient to consider the small difference
 \ba
 \xi\;\equiv\;\frac{\pi}{2}\;-\;|\epsilon|\;\;\;,
 \label{}
 \ea
 in terms whereof the inverse quality factor will read:
 \ba
 Q^{-1}\,=\,\frac{\cot \xi}{1\;-\;\xi
 \;\cot\xi}\;=\;\frac{1}{\tan\xi\;-\;\xi}\;=\;\frac{1}{
 z\;-\;\arctan z}\;=\;\frac{1}{\frac{\textstyle 1}{\textstyle 3}\;z^3\;\left[
 \,1\;-\;\frac{\textstyle 3}{\textstyle 5}\;z^2\,+\;O(z^4)\,\right]}
 \;\;\;,\;\;\;
 \label{108}
 \ea
 where $\;z\,\equiv\,\tan\xi\;$ and, accordingly,
 $\;
 \xi\;=\;\arctan z\;=\;z\,-\frac{\textstyle 1}{\textstyle 3}\,z^3\,+\,\frac{\textstyle 1}{\textstyle 5}\,z^5\,+\,O(z^7)\;\,.
 \;$
 Formula (\ref{108}) may, of course, be rewritten as
 \ba
 z^3\;\left[\,1\;-\;\frac{3}{5}\;z^2\;+\;O(z^4)\;\right]\;=\;3\;Q
 \;\;\;
 \label{}
 \ea
 or, the same, as
 \ba
 z\;=\;(3\,Q)^{1/3}\;\left[\,1\;+\;\frac{1}{5}\;z^2\;+\;O(z^4)\,\right]\;\;\;.
 \label{}
 \ea
 While the zeroth approximation is simply $\;z\,=\,(3Q)^{1/3}\,+\,O(Q)\;$, the first
 iteration gives:
 \ba
 \tan\xi\;\equiv\;z\;=\;
 (3Q)^{1/3}\,\left[\,1\;+\;\frac{1}{5}\;(3Q)^{2/3}\;+\;O(Q^{4/3})\,\right]\;
 =\;q\;\left[\,1\;+\;\frac{1}{5}\;q^2\;+\;O(q^4)\,\right]\;\;\;,~~~
 \label{}
 \ea
 with $\,q\,=\,(3Q)^{1/3}\,$ playing the role of a small parameter.

 We now see that the customary relation (\ref{customary}) should be substituted, for
 large lags, i.e., for small\footnote{~The afore-employed expansion of $\,\arctan z\,$ is valid for
 $\,|z|\,<\,1\,$. This inequality, along with (\ref{108}), entails: $\,Q\,=\,z\,-\,\arctan z\,<\,1\,-\,\pi/4\,$.}
 values of $\,Q\,$, with:
 \ba
 \tan|\epsilon|\;=\;(3Q)^{-1/3}\,\left[\,1\;-\;\frac{1}{5}\;(3Q)^{2/3}\;+\;O(Q^{4/3})\,\right]
 \label{}
 \ea
 The formula for the tidal torque contains a multiplier
 $\,\sin \epsilon\,\cos\epsilon\,$, whose absolute value can, for our purposes, be written down as
 \ba
 \sin|\epsilon|\,\cos|\epsilon|=\cos\xi\,\sin\xi=\frac{\tan\xi}{1+\tan^2\xi}
 \,=\,\frac{q\,\left[1+\frac{\textstyle 1}{\textstyle 5}\,q^2+O(q^4)\right]}{1+q^2\left[1
 +O(q^2)\right]}
  =(3Q)^{1/3}\left[1-\frac{4}{5}(3Q)^{2/3}+O(Q^{4/3})\right]\;,~~~
 \label{}
 \ea
 whence
 \ba
 \sin\epsilon\;\cos\epsilon\;=\;\pm\;(3Q)^{1/3}\left[1-\frac{4}{5}(3Q)^{2/3}+O(Q^{4/3})\right]\;,~~~
 \label{}
 \ea
 an expression vanishing for $\,Q\,\rightarrow\,0\;$. Be mindful that both $\,\epsilon_{\textstyle{_{2200}}}\,$
 and the appropriate component of the torque change their sign on the satellite crossing the synchronous
 orbit.

 \end{document}